\documentclass[11pt]{article} 

\usepackage{graphicx}
\usepackage{times}
\author{Norman Matloff}

\title{Software Alchemy:  Turning Complex Statistical Computations
into Embarrassingly-Parallel Ones}

\author{Norman Matloff \\
\\
Department of Computer Science \\
University of California, Davis}

\begin{document}

\maketitle

\begin{abstract}
{The growth in the use of computationally intensive statistical
procedures, especially with Big Data, has necessitated the usage of
parallel computation on diverse platforms such as multicore, GPU,
clusters and clouds.  However, slowdown due to interprocess
communication costs 
typically limits such methods to ``embarrassingly parallel'' (EP)
algorithms, especially on non-shared memory platforms.  This paper
develops a broadly-applicable  method for converting many non-EP
algorithms into statistically equivalent EP ones.  The method is shown
to yield excellent levels of speedup for a variety of statistical
computations.  It also overcomes certain problems of memory
limitations.} 

\end{abstract}

% \keywords{parallelization, embarrassingly parallel 
% algorithms, Big Data, cloud computing, GPU computing, memory limitations, 
% i.i.d.\ estimators, asymptototically normal estimators}

\section{Introduction}
\label{intro}

Many modern statistical methods involve computationally intensive
statistical algorithms, and are often applied to data sets with large
numbers of records and/or variables.  Indeed, in this era of Big Data,
it is common to have millions of records, but even with tens of
thousands of cases computation may present a real challenge.

This has necessitated the usage of parallel computational methods on
diverse platforms such as multicore, graphics processing units (GPUs),
clusters and clouds.  Interest in such methods is keen.  Consider the R
statistical language, for instance.  The central list of parallel
processing tools for R, CRAN Task View on High-Performance and Parallel
Computing with R, includes dozens of general and application-specific
packages.  Many of these packages did not exist at the time a broad
survey paper on the topic was published in 2009 \cite{schmid}, so the
growth in interest is remarkable.

However, slowdown due to interprocess communication costs (network
delays, memory consistency actions and so on) often limits such methods
to ``embarrassingly parallel'' (EP) algorithms.  Technically, the term
EP applies to algorithms that are so easy to parallelize that it is
``embarrassing,'' in the sense of presenting  no challenge to the
programmer.  However, the term as typically used imposes an additional
condition that there is very little communication overhead among the
processes.  This is key in statistical contexts, in which algorithms are
often not EP.

% As a result, an algorithm that is highly efficient in a single-thread
% implementation may work poorly when parallelized.  In the parallel
% setting, it may be faster overall to use an EP computation whose
% single-thread version is very inefficient.
% 
% For example, Garcia (2008) used a kd-tree algorithm for nearest-neighbor
% computation on a single processor, but resorted to the highly EP
% algorithm using ``brute force'' for the parallel version.  Efficient
% parallelization of kd-tree has been accomplished in some applications
% but in general is quite difficult, due to communication overhead
% (Shevtsov et al, 2007).  Thus it may be expedient to settle for an
% inefficient EP algorithm.
% 
% Sorting operations form the core of the computation of many statistical
% estimators, such as trimmed means, nonparametric estimators and so on.
% Although Quicksort is highly efficient sorting method in single-threaded
% contexts, it has generally been found that sample sort, which is
% essentially an EP algorithm, works better on parallel platforms.
% (Tsigas and Zhang did find on one specific platform that a highly
% ``tweaked'' version of Quicksort was superior to sample sort, but they
% did not optimize the latter.)

This paper provides a solution to the problem, applicable generally to
statistical estimators based on i.i.d.\ samples.  In essence, it converts
a non-EP algorithm to an EP algorithm of the same statistical accuracy.  
The method itself is easy to explain.  It begins with a traditional
method of parallel processing, in which one partitions the data into
chunks, applies an algorithm to each chunk, and then somehow combines
the results of the chunks.  That word {\it somehow} is the central issue
in traditional approaches; it is in this results-combining stage that
problems often arise, as the combining procedure typically involves
slow, non-EP computation.  

Consider {\it mergesort}, a method for sorting an array of numbers.
Here the array is broken into chunks, and each process sorts one of the
chunks (using a nonparallel algorithm).  The sorted chunks are then
merged, and it is there that the slowdown occurs:  The merging process
is not EP.  With this example in mind, we might take a working
definition of EP to be problems of {\it map-reduce} form (\cite{lee}),
providing that the reduce portion involves little or no computation.

What makes the method developed here different is that it exploits 
{\it statistical} properties.  The idea is to combine the chunks simply
by averaging them, an operation involving essentially no non-EP work,
and then recognize the statistical properties of this averaging.  This
last point will form the crux of this paper.

We will refer to this chunking and averaging approach as CA.  Clearly, a
core question must be answered:  CA would be of little value unless it
were verified that the proposed estimator has the same statistical accuracy
as the original, nonchunked one.  (We will refer to the latter as the {\it
full} estimator, FE.) If the CA method were to not produce a consistent
estimator, or if it were to produce larger standard errors than those of
the full estimator, the speedups would not hold much value.  

Thus the key to CA is showing that it produces the same asymptotic
standard errors as FE.  This property is indeed verified here, in Section
\ref{asympt}, for asymptotically normally distributed FEs.  In other
words, if the FE is asymptotically normal, then CA has the same
asymptotic standard errors as FE.  Since most widely-used estimators in
statistics are in this category, the CA method is broadly applicable.

CA is inherently EP, and thus achieves software alchemy, turning non-EP
algorithms into EP ones.  In other words, {\it CA provides a fairly
general method for parallelizing statistical computation.}

% For example, Seligman (2010) found that speedups accrued from applying
% a GPU to linear regression problems occurred only for extremely large
% problems, involving more than 1000 variables.  Our chunking method 
% attains speedups even for 50 variables.

% But again, this phenomenon is possible only if the chunking method gives
% us the same statistical accuracy.  This will be shown in Section
% \ref{asympt}.

Another use of CA will be to circumvent memory limitations.  Today's
very large data sets can exceed common memory size constraints.  Such
constraints can be either physical (insufficient RAM and swap space).
or software based, such as the R language's maximum
$2^{31}-1$ byte size for any single object.  As of this writing,
R is completing a transition process under which most objects
can now have size up to $2^{53}-1$, but even then some Big Data
applications' memory requirements may exceed the physical and virtual
memory of one's machine.  CA can remedy that problem.

\section{Related Work and Contribution of This Paper}

Though the general formulation and analysis presented here for CA is
new, some previous research has touched on the idea in focused specific
settings.  Following is an overview of previous work, and the
contribution of the present paper.

Chunking was proposed for speedup purposes in \cite{hegland}.  Here only
a specific estimator was considered, a certain additive nonparametric
regression model. 

Chunked estimation was then investigated in another special case, that
of linear regression, in \cite{fan}, motivated by memory requirement
problems rather than speed.  The authors allowed their chunk averaging
to be weighted, and they derived the optimal weights.  They also showed
that certain statistics would be $t$~distributed under the assumption of
normally-distributed populations, and studied the resulting chunked
estimators via simulation.  However, that work was restricted to linear
regression, not general types of estimators, and was not aimed at
parallel processing, i.e.\ speedup.

\cite{guha2009} developed a variant of CA, essentially what will be
called ``C without the A'' in Section \ref{nonpar}, with the goal of
speedup.  Some of the same authors did some finite-sample analysis of CA
in the linear regression case in \cite{guha2012}.

\cite{kleiner} developed a bootstrap approach related to CA, based on
applying the bootstrap method to a number of small random subsamples of
the original data.  Since any bootstrap method requires the user to
choose the values of ``hyperparameters''---size and numbers of
subsamples---the authors proposed an adaptive method to choose these
parameters.  They derived some asymptotic results, but also noted that
their method will fail in settings in which the ordinary bootstrap
fails.  On the other hand, they indicated how their method can be used
for time series data, something other CA methods have not yet been
extended to.

The question of parallel computation specifically for statistical
quantities was addressed in the context of database hardware and
software in \cite{cohen}.  There the emphasis was on parallelization of
sums computation, which could then be applied to some statistical
operations involving sums.

Though the names are similar, CA has little relation to {\it model
averaging} (\cite{claeskens}).  The latter method, and the related
technique of bagging \cite{brei}, are not intended as a mechanism for
parallel computation.

The present author's work on CA as ``software alchemy'' began in 2010
\cite{matlofffudan} as an effort to avoid the overhead of task queues in
shared-memory computing.  The contributions of this paper are as
follows:

\begin{itemize}

\item It is verified that CA works, i.e.\ is fully statistically efficient,
in fairly general settings.

\item A rough characterization is given of circumstances for which CA
produces a speedup. 

\item It is shown that CA can yield {\it superlinear} speedup, to a
degree rare in the parallel processing world.

\item It is shown that CA can bring a speedup even in the nonparallel
case.

\item Timing experiments are presented for a variety of statistical
methods.

\end{itemize}

\section{Statistical Properties of CA}
\label{asympt}

% Consider for example k-means clustering.  In a simple parallel algorithm
% for this computation, we might partition the original data into
% \lstinline{nprocs} chunks, where \lstinline{nprocs} is the number of
% processes.  Let c denote the desired number of clusters.  An outline of
% the textbf would be:
% 
% \begin{lstlisting}[numbers=left]
% master process sets initial values for the c centroids
% for iter = 1,...,niters
%    parallel for all processes prc = 1,...,p
%       set local sum[i] = 0, i = 1,...,c
%       for each observation in chunk prc
%          determine closest centroid cc to this observation
%          add this observation to local sum[cc]
%    master process computes new centroids
% \end{lstlisting}

Suppose we have i.i.d.\ data $V_i, ~ i = 1,...,n$ from some distribution
$F_V$.  Let $\theta$ denote some population value of interest, that is,
$\theta$ is some function of $F_V$.  The full estimator FE of $\theta$
based on $d$ observations is some function $b_d$ of those values.  So,
the full estimate of $\theta$ based on the first $n$ observations will be
denoted by

\begin{equation}
\label{full}
\widehat{\theta} = b_n(V_1,...,V_n)
\end{equation}

Note that $V_i$, $\theta$ and so on are possibly vector-valued.   

Partition the data into $r$ chunks, where $r$ is the number of parallel
processes; $r$ might be the number of cores in a multicore machine, for
example.  Assuming for now that $r$ evenly divides $n$, the chunk size
is $k = n/r$.\footnote{It would be natural to take the first chunk to
consist of $V_1,...,V_k$, the second one as $V_{k+1},...,V_{2k}$, and so
on.  Since the data are assumed i.i.d., any assignment of chunks would
work.  However, note the comments in Section \ref{noniidcase}.} Write
the j$^{th}$ observation in chunk $m$ as $W_{mj}$, $m = 1...r, j =
1...k$.  

Denote the estimator of $\theta$ on chunk $m$, i.e.\ FE applied to that
chunk, by $\widetilde{\theta}_{mr}$ (retaining the $r$ for clarity below).
Then we have

\begin{equation}
\widetilde{\theta}_{mr} = b_k(W_{m1},...,W_{mk})
\end{equation}

\noindent 
Now define the CA estimator:

\begin{equation}
\label{chunked}
\overline{\theta} = \frac{1}{r} 
\sum_{m=1}^r
\widetilde{\theta}_{mr}
\end{equation}

The key result will be that $\overline{\theta}$ is asymptotically
normal, and most important, that it has the same asymptotic covariance
matrix as $\widehat{\theta}$.  In other words, the statistical accuracy
of $\overline{\theta}$ is just as good as that of $\widehat{\theta}$.
This will now be shown.

It is assumed that the $V_i$ are independent and identically
distributed.  As will be discussed later, the latter assumption can be
dropped if the definition of the asymptotics is posed properly.

For notational convenience, the analysis here will treat the case of
scalar $\theta$.  The vector-valued case follows the same derivation
path.

It is assumed that $\widehat{\theta}$ has an asymptotic normal
distribution, i.e.\ that there exists some $\sigma$ for which

\begin{equation}
\label{anorm}
\lim_{n \rightarrow \infty} P 
\left [
\frac
{\sqrt{n}(\widehat{\theta} - \theta)} {\sigma}
\leq t
\right ] = \Phi(t)
\end{equation}

\noindent for all t, where $\Phi$ is the cdf for the N(0,1)
distribution.  It will shown below that $\overline{\theta}$ does just as
well as $\widehat{\theta}$, in the sense that for fixed r,
$\sqrt{rk}(\overline{\theta} - \theta)/\sigma$ converges in distribution
to N(0,1), as k goes to infinity.\footnote{It is clear that $k
\rightarrow \infty$ is generally also a necessary condition, as
otherwise $\overline{\theta}$ would fail to be even a consistent
estimator of $\theta$ in many applications.  The results here describe
the situation in which larger and larger applications are run on the
same r-core machine, r-node cluster etc.}

This can be shown in a variety of ways.  With a proper formulation, the
result could be shown to follow from the material in \cite{hartigan}.
But a simple derivation is as follows:

Let $g_k$ and $h_{rk}$ be the characteristic functions of $\sqrt{k}
(\widetilde{\theta}_{mr} - \theta)/\sigma$ and $\sqrt{rk}
(\overline{\theta}- \theta)/\sigma$, respectively; the former is the
characteristic function for the standardized estimator based on one
chunk, while the latter is the corresponding function for
$\overline{\theta}$ as a whole.  Then

\begin{eqnarray}
h_{rk}(t) &=&
E \left [
e^{it \sqrt{rk} (\overline{\theta} - \theta)/\sigma}
\right ] \\ 
&=& 
E \left [
e^{it \sqrt{rk} \cdot \frac{1}{r}  
   \sum_{m=1}^r (\widetilde{\theta}_{mr} - \theta)/\sigma}
\right ] \\ 
&=& 
\Pi_{m=1}^r
E \left [
e^{it \sqrt{k/r} (\widetilde{\theta}_{mr} - \theta)/\sigma}
\right ] \\ 
&=& 
\Pi_{m=1}^r
g_k(t \sqrt{1/r}) \\
&=& 
\left [
g_k(t \sqrt{1/r})
\right ]^r
\label{hrkeqn}
\end{eqnarray}

Under our normality assumption for $\widehat{\theta}$, which applies to
the $\widetilde{\theta}_{mr}$ as well since they are ``mini-versions'' of
$\widehat{\theta}$, we have that 

\begin{equation}
\label{limg}
\lim_{k \rightarrow \infty} g_{k}(t) = e^{-0.5 t^2}
\end{equation}

\noindent 
which is the characteristic function of the standard normal distribution.
Thus for fixed r,

\begin{equation}
\label{limk}
\lim_{k \rightarrow \infty} h_{rk}(t) 
= [ e^{-0.5 (t/\sqrt{r})^2} ]^r
= e^{-0.5 t^2}
\end{equation}

Thus $\overline{\theta}$ is asymptotically normal, and its asymptotic
variance $\sigma^2$ does match that of $\widehat{\theta}$, as promised.

% With some fairly technical work, one could go further, and show that the
% asymptotic convergence is uniform in r.  In other words, the convergence
% is equally fast on, say, smaller and larger multicore machines.  The
% proof would involve expanding $g_k(t/\sqrt{r})$ in a second-degree
% Taylor approximation and then using the well-known fact that
% 
% \begin{equation}
% \lim_{u \rightarrow \infty} (1+1/u)^u = e^u
% \end{equation}

Since CA will be employed almost exclusively on large data sets
anyway (small ones would not need the speed CA provides), the asymptotic
results should hold well in practice.  This is confirmed in the
empirical results presented in Section \ref{empirical} below.

\section{Refinements}

One can obtain standard errors for $\overline{\theta}$ empirically
(similar to bootstrapping).  The estimated covariance matrix of
$\overline{\theta}$ is

\begin{equation}
\frac{1}{r}
\sum_{m=1}^r 
( \widetilde{\theta}_{mr} - \overline{\theta} )
( \widetilde{\theta}_{mr} - \overline{\theta} )^\top
\end{equation}

\noindent
One can then obtain standard errors for the components of
$\overline{\theta}$ (and for linear combinations of them) in the 
usual manner.

Often $n$ will not be an exact multiple of $r$.  In such cases, let $k =
\lfloor n/r \rfloor$, and $k' = n - (r-1) k$.  For $i = 1,...,r-1$, let
chunk $i$ consist of $k$ observations as before, but define the $r^{th}$
chunk to consist of the last $k'$ observations.  

The weightings in Equation (\ref{chunked}) will no longer be $1/r$.  To
determine the proper weightings, note first that due to symmetry, the
weights for the first $r-1$ estimators will be the same, with a common
value we'll call $\lambda$.  The weight for the $r^{th}$ estimator will
then be $1-(r-1)\lambda$.  

% It is easily shown that for independent random variables A and B, the
% value of $\alpha$ that minimizes $Var[\alpha A + (1-\alpha) B]$ is
% 
% \begin{equation}
% \alpha = 
% \frac{Var(B)}{Var(A)+Var(B)}
% \end{equation}
% 
% Then set
% 
% \begin{equation}
% A =  \sum_{m=1}^{r-1}
% \widetilde{\theta}_{mr}
% \end{equation}
% 
% and 
% 
% \begin{equation}
% B =  \widetilde{\theta}_{rr}
% \end{equation}
% 
% with $\lambda$ playing the role of $\alpha$.
% 
% The asymptotic variances of A and B are proportional to (r-1)/k and 1/k',
% respectively.  The best value of $\lambda$ is then

To determine $\lambda$ , again think of the case of scalar $\theta$ for
convenience.  The asymptotic variance of the overall weighted estimator
$\overline{\theta}$ will be proportional to

\begin{equation}
\lambda^2 (r-1) \cdot \frac{1}{k} +
[1-(r-1)\lambda]^2 \cdot \frac{1}{k'} 
\end{equation}

Minimizing this with respect to $\lambda$, we find

\begin{equation}
\lambda = \frac
{1}
{r-1 + \frac{k'}{k}}
\end{equation}

The definition of the CA estimator then becomes a modified version of
Equation (\ref{chunked}):

\begin{equation}
\overline{\theta} = 
\sum_{m=1}^{r-1}
\lambda \widetilde{\theta}_{mk}
+ [1-(r-1)\lambda] \widetilde{\theta}_{rk}
\end{equation}

Sample R textbf for CA computation is presented in the Appendix.

\section{On What Types of Statistical Methods Will CA Be Faster?}
\label{alg}

The above findings indicate that CA will produce speedup in many types
of statistical methods.  But which types?

\subsection{Algorithmic Analysis}

Let us consider speedup from an algorithmic time complexity point of
view.  As before, let $n$ denote the number of observations in an
i.i.d.\  sample, with $r$ denoting the number of processes that work in
parallel.

Consider statistical methods needing time $O(n^c)$.  If we have $r$
processes, then CA assigns $n/r$ data points to each process.  Speaking
in rough, exploratory terms, CA would reduce the run time to $O((n/r)^c)
= O(n^c/r^c)$ time under CA, a speedup of $r^c$.  The larger the
exponent c is, the greater the speedup.

Thus statistical applications with computational time complexity greater
than or equal to $O(n)$ will be major beneficiaries of CA.  Examples of
such applications are quantile regression, estimating hazard functions
with censored data, and so on.  

The situation, though, is more subtle when there are two variables
affecting run time, not just $n$.  Look, for example, at linear regression
with $p$ predictor variables.  The time needed is first $O(np^2)$ to
form sums of squares and cross products, followed by $O(p^3)$ to derive
the least-squares estimates from those sums.  The latter, which is the
time complexity for matrix inversion or equivalent
operation,\footnote{If QR factorization is used, the complexity may be
$O(p^2)$, depending on exactly what is being computed.} is independent
of $n$ and thus not helped by CA.  Similar reasoning applies to
principal components analysis.  There the first phase consists of a
computation of sums that benefits from CA, but the second phase
involves eigenanalysis with time independent of $n$.  In both of these
examples, CA is helpful for the first stage of computation, so it is
helpful overall, but often only to a modest degree.

\subsection{Superlinear Behavior}

Remarkably, the CA method can enable performance increases on a {\it
superlinear} scale.  This term, from the parallel processing literature,
refers to speedups of more than $r$ from only $r$ processes.  

In the general parallel processing world, this phenomenon is quite rare.
Moreover, when it does occur, it is small in size, and arises from
ancillary cache effects and the like; see for instance \cite{kosec}.
But in the statistical world, users of CA will find superlinear speedups
to be commonplace and quite substantial, as follows.

As noted above, CA reduces the time complexity of an $O(n^c)$ problem to
roughly $O(n^c/r^c)$ for a statistically equivalent problem, whereas a
linear speedup would only reduce the time to $O(n^c/r)$.  If $c > 1$,
then the speedup obtained from CA is greater than linear in r, hence the
term ``superlinear.''  

Indeed, in the superlinear case, CA may yield a speedup even on a
uniprocessor machine.  Here is a back-of-the-envelope computation: The
$r$ chunks must now be computed serially rather than in parallel, and
take time $r ~ O(\frac{n^c}{r^c}) = O(\frac{n^c}{r^{c-1}})$.  This
suggests that for $c > 1$, CA may be faster than the FE even in
uniprocessor settings.  The same reasoning shows that it may pay to {\it
oversubscribe} one's hardware, e.g.\ have more threads than cores on a
multicore machine.

\section{Empirical Investigation}
\label{empirical}

To illustrate the value of CA, the method was used on the following
diverse set of statistical applications:

\begin{itemize}

\item Kendall's $\tau$ correlation ($O(n^2)$ and
$O(n \log n)$ algorithms),

\item quantile regression,

\item hazard function estimation, 

\item log-concave density function estimation, and 

\item linear regression. 

\end{itemize}

% Except for the last case, the time complexity of the statistical
% method grows faster than n, and thus CA should attain a superlinear
% speedup.  We would expect CA to have mixed performance in the last 
% case.

Simulation runs were conducted on an Intel Xeon 2.0 Gz 64-bit machine
with 32 cores and a hyperthreading degree of 2.  Thus as many as 64
threads may be in computation at once.  Runs with 4, 8, 16, 24, 32 and
40 threads were conducted.

In most cases the problem size was chosen to be large enough for
parallelization to be worthwhile.  The criterion for the latter was
arbitrarily set to having an FE run time of at least a few seconds.

The R language was used, using R's standard
packages for each of the statistical methods listed above.
Interprocess communication used R's \textbf{parallel} package,
in the portion derived from the \textbf{snow} package.  The algorithm can
easily be ported to R packages that run under
Hadoop, such as \textbf{rmr} or \textbf{RHIPE}.

Elapsed times were recorded in seconds.  As will be seen, CA yielded
speedups in all cases, but as predicted in Section \ref{alg}, the
magnitude of speedup varied widely.  In some cases, the speedup was
superlinear, while in others it was much more modest.  Where feasible, a
45-degree line was also plotted, to asses possible superlinearity.

In addition to measuring run times, the $l_1$ relative absolute
differences between CA ($\overline{\theta}$) and the original estimator
($\widehat{\theta}$) were computed.  For $p$-component $\theta$, this is

\begin{equation}
\frac
{\sum_{i=1}^p |\overline{\theta}_i - \widehat{\theta}_i|} 
{\sum_{i=1}^p |\widehat{\theta}_i| }
\end{equation}

\noindent 
Values are reported to four decimal places.  In all cases this 
relative difference was negligible, amplifying the point that the two 
estimators have the same asymptotic distribution.

\subsection{Results}

\bigskip

{\bf Kendall's $\tau$:}

\bigskip

Here (X,Y) pairs were formed, as follows.  Independent U(0,1)
variates $U_1$ and $U_2$ were generated, and then (X,Y) was formed as 

\begin{equation}
(X,Y) = (U_1,0.2U_1+U_2) 
\end{equation}

\begin{figure}[!t]
\centering
\includegraphics[width=3.5in]{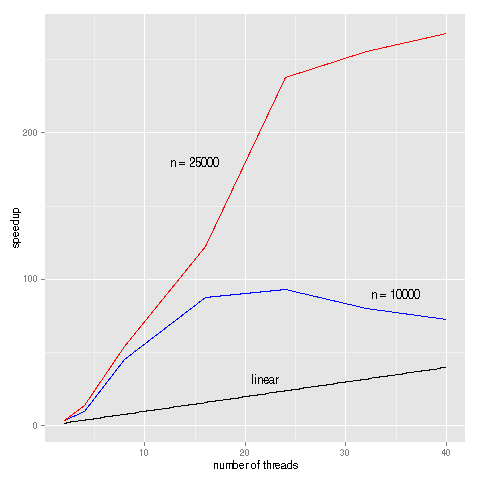}
\caption{Kendall's $\tau$}
\label{ktau}
\end{figure}

The R function \textbf{cor.test()} was used, with the argument
\textbf{method = "kendall"}.  The results are shown in Figure \ref{ktau}.
% The results are shown in Tables \ref{kend1} and \ref{kend2}.
Here the FE times were 2.98 and 17.13, and relative difference value
ranges were 0.0010-0.0093 and from 0.0000-0.0020, respectively for $n$ =
10000 and 25000.

% \begin{table}
% \begin{center}
% \begin{tabular}{|r|r|r|r|r|}
% \hline
% thrds & speedup & rel diff \\ \hline
% 2 & 3.92 & 0.0010 \\ \hline 
% 4 & 9.63 & 0.0013 \\ \hline 
% 8 & 45.08 & 0.0013 \\ \hline 
% 16 & 87.50 & 0.0058 \\ \hline 
% 24 & 92.97 & 0.0037 \\ \hline 
% 32 & 80.41 & 0.0093 \\ \hline 
% 40 & 72.56 & 0.0070 \\ \hline 
% \end{tabular}
% \end{center}
% \caption{Kendall, n = 10000, FE time 2.98s.}
% \label{kend1}
% \end{table}
% 
% \begin{table}
% \begin{center}
% \begin{tabular}{|r|r|r|r|r|}
% \hline
% thrds & speedup & rel diff \\ \hline
% 2 & 3.63 & 0.0000 \\ \hline 
% 4 & 14.01 & 0.0003 \\ \hline 
% 8 & 54.55 & 0.0000 \\ \hline 
% 16 & 122.36 & 0.0020 \\ \hline 
% 24 & 237.92 & 0.0012 \\ \hline 
% 32 & 255.67 & 0.0005 \\ \hline 
% 40 & 267.65 & 0.0009 \\ \hline 
% \end{tabular}
% \end{center}
% \caption{Kendall, n = 25000, FE time 17.13s.}
% \label{kend2}
% \end{table}

The ``ordinary'' Kendall algorithm has an $O(n^2)$ time complexity.
Section \ref{alg} indicates that we should expect superlinear behavior,
which is confirmed here, especially for the larger value of n.  The
timings here suggest similar performance for other methods of this type,
such as any U-statistic.

\begin{figure}[!t]
\centering
\includegraphics[width=3.5in]{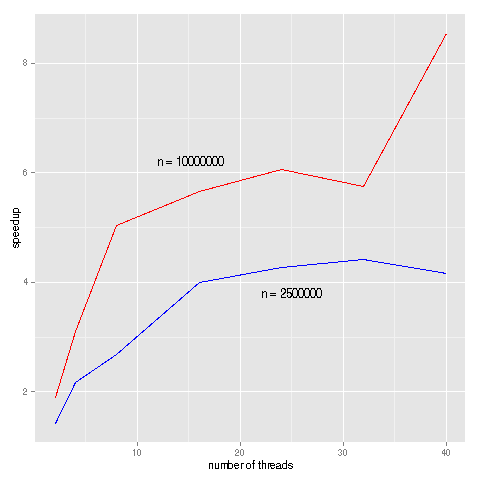}
\caption{Kendall's $\tau$, Knight's algorithm}
\label{ktaufast}
\end{figure}

But the clever algorithm due to Knight (\cite{christensen}) has only
$O(n \log n)$ time complexity.  This is implemented in the function
\textbf{cor.fk()} of the R package \textbf{pcaPP}.  CA is beneficial
here too, though in much larger problems and then only modestly compared
to the number of threads used; see Figure \ref{ktaufast}.

Here the FE times were 3.24 and 22.05, and relative difference values
were zero to four decimal places, for both $n$ =
2500000 and 10000000.  No superlinear effect was observed at these
values of $n$, but speedups of up to a factor of more than 8 occurred.

% \begin{table}
% \begin{center}
% \begin{tabular}{|r|r|r|r|r|}
% \hline
% thrds & speedup & rel diff \\ \hline
% 2 & 1.43 & 0.0000 \\ \hline 
% 4 & 2.17 & 0.0000 \\ \hline 
% 8 & 2.68 & 0.0000 \\ \hline 
% 16 & 4.00 & 0.0000 \\ \hline 
% 24 & 4.28 & 0.0000 \\ \hline 
% 32 & 4.41 & 0.0000 \\ \hline 
% 40 & 4.17 & 0.0000 \\ \hline 
% \end{tabular}
% \end{center}
% \caption{Fast Kendall, n = 2500000, FE time 3.24s.}
% \label{knight1}
% \end{table}
% 
% \begin{table}
% \begin{center}
% \begin{tabular}{|r|r|r|r|r|}
% \hline
% thrds & speedup & rel diff \\ \hline
% 2 & 1.90 & 0.0000 \\ \hline 
% 4 & 3.11 & 0.0000 \\ \hline 
% 8 & 5.04 & 0.0000 \\ \hline 
% 16 & 5.66 & 0.0000 \\ \hline 
% 24 & 6.06 & 0.0000 \\ \hline 
% 32 & 5.75 & 0.0000 \\ \hline 
% 40 & 8.54 & 0.0000 \\ \hline 
% \end{tabular}
% \end{center}
% \caption{Fast Kendall, n = 10000000, FE time 22.05.}
% \label{knight2}
% \end{table}

{\bf Quantile regression:}

\begin{figure}[!t]
\centering
\includegraphics[width=3.5in]{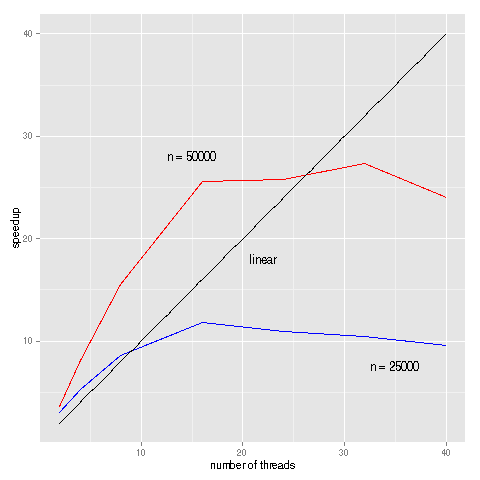}
\caption{Quantile Regression}
\label{quant}
\end{figure}

Here $p$ is the number of predictor variables, and each observation was
of the form $(X_1,...,X_p,Y)$, with the $X_i$ being i.i.d.\ U(0,1) and
with

\begin{equation}
Y = X_1+...+X_p + 0.2 U
\end{equation}

\noindent
where U had a U(0,1) distribution and was independent of the
$X_i$.  The function \textbf{rq()} from the package \textbf{quantreg} 
was used, with $p$ = 75.  The results are displayed in 
Figure \ref{quant}.

Here the FE times were 16.20 and 60.89, and relative difference value
ranges were 0.0000-0.0010 and from 0.0003-0.0009, respectively for $n$ =
25000 and 50000.  Strong speedups are obtained, sometimes superlinear.

% \begin{table}
% \begin{center}
% \begin{tabular}{|r|r|r|r|r|}
% \hline
% thrds & speedup & rel diff \\ \hline
% 2 & 3.01 & 0.0005 \\ \hline 
% 4 & 5.31 & 0.0007 \\ \hline 
% 8 & 8.59 & 0.0010 \\ \hline 
% 16 & 11.86 & 0.0001 \\ \hline 
% 24 & 10.90 & 0.0000 \\ \hline 
% 32 & 10.41 & 0.0000 \\ \hline 
% 40 & 9.58 & 0.0000 \\ \hline 
% \end{tabular}
% \end{center}
% \caption{Quantile, n = 25000, p = 75, FE time 16.20s.}
% \label{q1}
% \end{table}
% 
% \begin{table}
% \begin{center}
% \begin{tabular}{|r|r|r|r|r|}
% \hline
% thrds & speedup & rel diff \\ \hline
% 2 & 3.58 & 0.0003 \\ \hline 
% 4 & 8.06 & 0.0004 \\ \hline 
% 8 & 15.51 & 0.0005 \\ \hline 
% 16 & 25.55 & 0.0006 \\ \hline 
% 24 & 25.81 & 0.0006 \\ \hline 
% 32 & 27.34 & 0.0008 \\ \hline 
% 40 & 24.07 & 0.0009 \\ \hline 
% \end{tabular}
% \end{center}
% \caption{Quantile, n = 50000, p = 75, FE time 60.89s.}
% \label{q2}
% \end{table}

\bigskip

{\bf Hazard function estimation:}

\bigskip

\begin{figure}[!t]
\centering
\includegraphics[width=3.5in]{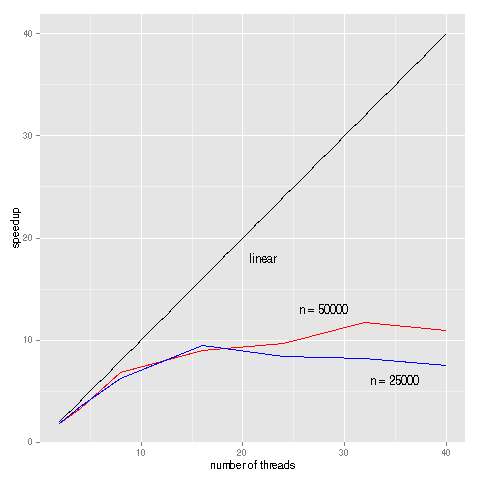}
\caption{Hazard Function Estimation}
\label{haz}
\end{figure}

Here $p$ is the proportion of censored observations.  The data were
sampled from U(0,1), and the parameter vector $\theta$ consisted of the
values of the hazard function h(t) at $t$ = 0.2, 0.4, 0.6, 0.8.  The
function \textbf{muhaz()} from the CRAN package of the same name 
was used, with the default settings.
The results are shown in Figure \ref{haz}.

Speedups here were well below linear (except for small numbers of
threads), but were still strong, as large as 10 or more.
FE times were 5.59 and 10.97, and relative difference value ranges were
0.0031-0.0106 and from 0.0002-0.0108, respectively for $n$ = 25000 and
50000.

% \begin{table}
% \begin{center}
% \begin{tabular}{|r|r|r|r|r|}
% \hline
% thrds & speedup & rel diff \\ \hline
% 2 & 1.87 & 0.0017 \\ \hline 
% 4 & 3.25 & 0.0055 \\ \hline 
% 8 & 6.86 & 0.0074 \\ \hline 
% 16 & 9.02 & 0.0100 \\ \hline 
% 24 & 9.68 & 0.0091 \\ \hline 
% 32 & 11.69 & 0.0079 \\ \hline 
% 40 & 10.97 & 0.0108 \\ \hline 
% \end{tabular}
% \end{center}
% \caption{Hazard, n = 50000, p = 0.2, FE time 10.97.}
% \label{haz}
% \end{table}

{\bf Log concave density estimation:}

\begin{figure}[!t]
\centering
\includegraphics[width=3.5in]{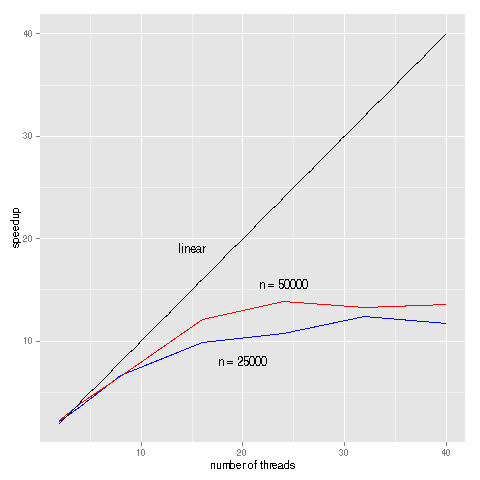}
\caption{Log-Concave Density  Estimation}
\label{lcd}
\end{figure}

This is a type of nonparametric density estimation.  The
data were generated from N(0,1), and $\theta$ was taken to be the value
of the density at 0.  The function \textbf{logConDens()} from the CRAN
package \textbf{logcondens} was used.  See Figure \ref{lcd} for the
results.  

The pattern here is similar to that found for hazard function
estimation, near-linear only for smaller number of threads.  FE times
were 28.90 and 67.92, and relative difference value ranges were
0.0031-0.0105 and zero to four decimal places, respectively for $n$ =
25000 and 50000.

% \begin{table}
% \begin{center}
% \begin{tabular}{|r|r|r|r|r|}
% \hline
% thrds & speedup & rel diff \\ \hline
% 2 & 2.17 & 0.0117 \\ \hline 
% 4 & 3.60 & 0.0014 \\ \hline 
% 8 & 6.63 & 0.0117 \\ \hline 
% 16 & 9.84 & 0.0004 \\ \hline 
% 24 & 10.72 & 0.0105 \\ \hline 
% 32 & 12.43 & 0.0094 \\ \hline 
% 40 & 11.75 & 0.0030 \\ \hline 
% \end{tabular}
% \end{center}
% \caption{Log concave, n = 25000, FE time 28.90s.}
% \label{lc}
% \end{table}

{\bf Linear regression:}

\begin{figure}[!t]
\centering
\includegraphics[width=3.5in]{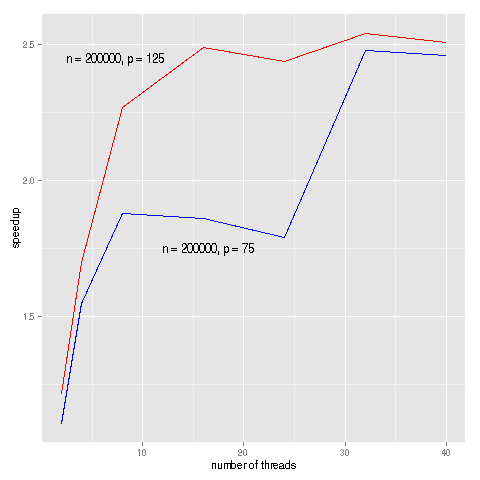}
\caption{Linear Regression}
\label{lm}
\end{figure}

Here $p$ is the number of predictor variables, and the distribution of
the data was as in the quantile regression case above.  The outcomes are
shown in Figure \ref{lm}.  FE times were 5.22 and 11.06, and relative
difference values were zero to four decimal places, for both $p$ = 75
and 125.  For the reasons explained in Section \ref{alg}, prospects for
speedup in the linear regression case are somewhat limited, but still CA
yields approximately 50\% speedup even with only 4 cores.

% \begin{table}
% \begin{center}
% \begin{tabular}{|r|r|r|r|r|}
% \hline
% thrds & speedup & rel diff \\ \hline
% 2 & 1.11 & 0.0000 \\ \hline 
% 4 & 1.55 & 0.0000 \\ \hline 
% 8 & 1.88 & 0.0000 \\ \hline 
% 16 & 1.86 & 0.0000 \\ \hline 
% 24 & 1.79 & 0.0000 \\ \hline 
% 32 & 2.48 & 0.0000 \\ \hline 
% 40 & 2.46 & 0.0000 \\ \hline 
% \end{tabular}
% \end{center}
% \caption{Linear, n = 200000, p = 75, FE time 5.22s.}
% \label{lin1}
% \end{table}
% 
% \begin{table}
% \begin{center}
% \begin{tabular}{|r|r|r|r|r|}
% \hline
% thrds & speedup & rel diff \\ \hline
% 2 & 1.22 & 0.0000 \\ \hline 
% 4 & 1.70 & 0.0000 \\ \hline 
% 8 & 2.27 & 0.0000 \\ \hline 
% 16 & 2.49 & 0.0000 \\ \hline 
% 24 & 2.44 & 0.0000 \\ \hline 
% 32 & 2.54 & 0.0000 \\ \hline 
% 40 & 2.51 & 0.0000 \\ \hline 
% \end{tabular}
% \end{center}
% \caption{Linear, n = 200000, p = 125, FE time 11.06s.}
% \label{lin2}
% \end{table}

\subsection{Discussion}

In the above experiments,  CA achieved good performance in a diverse set
of statistical methodologies, with significant speedups even in the more
modest cases.  In addition, CA achieved this in a \underline{convenient}
manner---the straightforward, simple R routine in the
Appendix was used for all the applications.

The CA method is based on asymptotics.  As pointed out earlier, since
the performance boost of parallel processing is generally needed only on
large samples, the asymptotics should work well.  This was confirmed in
the simulations, which showed that the relative difference between CA
and FE was negligible even for just moderately-large data sets.  

In some cases, speedup decreased slightly at the high end of numbers of
threads.  Though this may be due to sampling variation, it should be
noted that although each of the threads is doing the same amount of
work,\footnote{If the number of processes evenly divides $n$.} there may
be some variation from one thread to another; with more threads, the
maximum interthread variation will increase.  

% As expected, the improvement in the linear regression case was much more
% modest than in the other examples, with the speedup factor reaching only
% about 2.50 even with 32 or more threads.  Note that experiments reported
% for GPU computation of linear models find that even to ``break even,''
% i.e. achieve a factor of 1.00 one needs 1000 variables, compared to the
% speedup of 2.50 with 75 variables here.

If the basic estimator (FE) itself is unstable and of questionable use,
CA may produce different results from FE.  This may occur, for instance,
in hazard function estimation when the proportion of censored
observations is high.  CA may actually have a stabilizing effect in such
cases, say if CA is used with medians instead of means.  This is a
topic for future study, possibly in connection to the literature on
model averaging.

% \section{A Preliminary Look at the Uniprocessor Case}
% 
% It was remarked earlier that a rough time complexity analysis suggests
% that CA may yield a speedup for some statistical estimators even in
% uniprocessor settings.  Since the focus of the present paper is on
% parallel processing, an in-depth investigation of this phenomenon is
% deferred to future work.  Here, though, is a preliminary look.
% 
% Simulations were conducted in the same settings as in Section
% \ref{empirical}, except that only a single thread on a single machine.
% The r chunks were run sequentially, one after another.  Figure
% \ref{nonparspeed} charts the speedups obtained for the various estimators.
% 
% \begin{figure*}
% \includegraphics[width=4.5in]{UniRatios.pdf}
% \label{nonparspeed}
% \caption{Nonparallel speedups}
% \end{figure*}
% 
% As to be expected, overhead issues limit the amount of speedup, but
% still there is good performance in the case of quantile regression and
% Kendall's $\tau$, and a 20\% speedup is attainable for log-concave
% density estimation.
% 
% The case of hazard function estimation is interesting.  Though CA brings
% good speedup in the parallel case, it yields no improvement in
% nonparallel settings.  Note, however, that this was using the default
% arguments for the {\tt muhaz()} function, in which kernel smoothing is
% performed; I found that simulations (not included here) of the
% nearest-neighbor version, specified via the argument {\tt
% bw.method="knn"}, showed very strong speedups using CA, even in
% nonparallel settings.

\section{``C without the A''}
\label{nonpar}

\cite{guha2009} were interested in visualization of very large datasets,
Their solution to the huge computation involved was to use a form of CA,
which they termed {\it Divide and Recombine} (D\&R).  The `R' here
roughly corresponds to the `A' in ``CA'' in the present paper.  In that
setting, though, there was no `A' and even the `R' meant only collecting
the chunk results into a single database of graphs.  As mentioned
earlier, they subsequently did bring in the `A' for the linear
regression setting in \cite{guha2012} (and alluded to ``numerical''
applications in \cite{xi}).  

But the emphasis in this section is on ``C without the A,'' a very
useful variant of CA that was essentially the strategy used in
\cite{guha2009}.  We will call it CWA, meaning that we divide into
chunks and apply some estimator to each chunk, but then somehow use the
resulting collection of estimators separately rather than, say,
averaging them.  

The main example here will involve density estimation.  Suppose we wish
to compute a nonparametric density estimate at each observations in our
sample.  Instead of using the entire data set for this at any given
point, we use only the data in the chunk to which the point belongs.

Why is CWA so computationally advantageous?  The results of Section
\ref{alg} provide an explanation.  Computation for a graph is at least
linear in n, the number of observations, and can be worse, say if
smoothing is involved.  Consider a kernel-based density estimator, for
instance.  If we graph the estimated density at each data point, we need
to do $O(n^2)$ work.

The significance of the results in Section \ref{alg} in the present
context is as follows.  The amount of computation needed is growing as
$O(n^2)$, yet the amount of statistical value we get is growing only as
$O(n)$, e.g.\ in variance of estimators.  So, the extra work involved
with doing density estimation on the full data set will not produce a
commensurate improvement in statistical quality.  

Moreover, in the case of graphics, in which small differences may not
adversely affect our visual perception, the large $n$ may not be very
useful anyway.  This consideration becomes even more important in light
of the fact that the nonzero nature of pixel width on the screen makes
the use of larger data sets meaningless at some point, due to
overplotting and the ``black screen effect.''  So CWA won't be harmful
in large data, and will yield large savings in run time.  In other
words, CWA is a win.

The author also used CWA to great benefit in \cite{matloffjsm},
which presented several novel visualization techniques for large
datasets.  (Here ``large'' simply meant large enough for graphing of the
full data to produce significant overplotting, which can occur even for
moderately-large values of $n$.)  For example, one of the methods proposed
was a novel approach to the screen-clutter problem in plotting parallel
coordinates.  The author used k-Nearest Neighbor methods to estimate
multivariate density. as had been done with parallel coordinates
previously, but with the new twist that only the lines with highest
density are plotted.  To compute the estimated density at the $i^{th}$
observation $V_i$, the author only used the data in chunk $j$, i.e.\
$V_{(j-1)k+1},...V_{jk}$.

% Since only a very small number of lines are
% plotted, there is no screen clutter, while those lines have a
% ``typicalness'' property (due to having maximal density) that may
% provide insight into relations between the variables.

This topic of CWA will be pursued further in future research.

\section{The Case of Non-i.i.d. Data} 
\label{noniidcase}

The derivation in Section \ref{asympt} assumed i.i.d.\ data.  This of
course is a standard condition in many statistical methods, and many
packages in both base R and CRAN assume it.  In our context
here of CA methods, \cite{kleiner} also assumes i.i.d.\ throughout,
except for the final section, which briefly discusses an extension to
time series.

Our derivation here could be generalized to non-identically distributed
data.  This would involve a proper definition of the term {\it
asymptotic}, along the lines of the classic analysis in \cite{jennrich},
in which it is essentially assumed that the empirical cdf of the data
converges to some distribution.

% In any case, a reasonable remedy in the nonidentically distributed case
% is to simply randomize the observations before performing the analysis.
% For instance, the data storage may be ordered by some categorical
% variable, say state or province.  Randomization would be useful here.

\section{Conclusions and Future Work}
\label{future}

The method developed here turns nonembarrassingly parallel calculations
into statistically equivalent embarrassingly parallel ones.  It was
found here to produce excellent speedups in a diverse set of
applications. It is also very easy and convenient to use, requiring no
expertise in parallel algorithms or subsampling techniques.

As noted earlier, areas for future investigation are the study of
possible stability-enhancing effects of CA for iterative algorithms, a
formal derivation for the nonidentically distributed case, and further
investigation into CWA.  Extension to time series models would also be
of interest.

\section{Acknowledgements}

The work has benefited from conversations with Michael Kane. 

\appendix

\section{R Code for CA Computation}

One of the virtues of CA is its simplicity, as seen in the following
R \textbf{snow} textbf.\footnote{Intended for illustration
purposes only.  Optimal implementation of CA depends on the platform.}
Here \textbf{cls} is the  \textbf{snow} cluster; \textbf{z} is the data matrix, one
observation per row; \textbf{probpars} is an R list, containing
possible further information to be made available to \textbf{sf}; and
\textbf{sf} is the user-supplied function to calculate the FE estimate.
The return value is the CA estimate.

\begin{verbatim}
# chunks averaging method, implemented for R's 'parallel' (formerly
# Snow) package

# arguments:
 
#    cls: 'parallel' cluster
#    z:  data (data.frame, matrix or vector), one observation per row
#    ovf:  overall statistical function, say glm()
#    estf:  function to extract the point estimate (possibly
#           vector-valued) from the output of ovf()
#    estcovf:  if provided, function to extract the estimated variance
#          f   of covariance matrix of the output of estf()
 
# value:
# 
#    R list, consisting of the CA-computed point estimate, and ,
#    optionally the estimate covariance matrix

ca <- function(cls,z,ovf,estf,estcovf=NULL) {
   require(parallel)
   if (is.data.frame(z)) z <- as.matrix(z)
   if (is.vector(z)) z <- matrix(z,ncol=1)
   n <- nrow(z)
   rowchunks <- clusterSplit(cls,1:n)
   chunks <- lapply(rowchunks,function(rowchunk) z[rowchunk,])
   ni <- sapply(rowchunks,length)  # chunk sizes
   wts <- ni / n  # weights in the averaging
   ovout <- clusterApply(cls,chunks,ovf)
   thts <- lapply(ovout,estf)
   lth <- length(thts[[1]])
   tht <- rep(0.0,lth)
   if (!is.null(estcovf)) {
      thtcov <- matrix(0,nrow=lth,ncol=lth)
      thtcovs <- lapply(ovout,estcovf)
   }
   for (i in 1:length(thts)) {
      wti <- wts[i]
      tht <- tht + wti * thts[[i]]
      if (!is.null(estcovf)) 
         thtcov <- thtcov + wti^2 * thtcovs[[i]]
   }
   res <- list()
   res$tht <- tht
   if (!is.null(estcovf)) res$thtcov <- thtcov
   res
}

\end{verbatim}

% \bibliographystyle{asa}
% \bibliography{ArXiv}

\end{document}